\documentclass[10pt, onecolumn]{article}

\usepackage[dvips]{graphicx}

\usepackage{amssymb,amsfonts,amsmath}

\begin{document}



\begin{titlepage}
\begin{center}
\Huge{Fluctuation-dissipation relations far from equilibrium}

\normalsize

 Jianhua Xing  \\
Department of biological sciences, Virginia Polytechnic Institute and State University, Blacksburg, VA 24061

\end{center}




\begin{abstract} 
The fluctuation-dissipation (F-D) theorem is a fundamental result for systems near thermodynamic equilibrium, and justifies studies between microscopic and macroscopic properties. It states that the nonequilibrium relaxation dynamics is related to the spontaneous fluctuation at equilibrium. Most processes in Nature are out of equilibrium, for which we have limited theory. Common wisdom believes the F-D theorem is violated in general for systems far from equilibrium. Recently we show that dynamics of a dissipative system described by stochastic differential equations can be mapped to that of a thermostated Hamiltonian system, with a nonequilibrium steady state of the former corresponding to the equilibrium state of the latter. Her we derived the corresponding F-D theorem, and tested with several examples. We suggest further studies exploiting the analogy between a general dissipative system appearing in various science branches and a Hamiltonian system. Especially we discussed the implications of this work on biological network studies.
\end{abstract}
\end{titlepage}





It is ubiquitous to observe a system at a state invariant with time (with the approximation that the relevant constraining parameters changes much slower than the time scale under interest). It can be a thermodynamic equilibrium state, or more likely a nonequilibrium steady-state. Some examples are  homeostatic states of living organisms, a stable eco- or financial system. Quite often it is important to know how a system initially subject to a perturbation relaxes to a steady state  (including the equilibrium state) after removal of the perturbation. The fluctuation-dissipation theorem states that the relaxation dynamics for a process close to equilibrium is related to the spontaneous fluctuation at equilibrium.  Originally formulated by Nyquist in 1928 \cite{Nyquist1928}, and first proved by Callen and Welton in 1951 \cite{Callen1951}, the F-D theorem is related to many important results in statistical physics. Examples are the Einstein-Smoluchowski relation on the diffusion constant and drag coefficient \cite{Einstein1905, Smoluchowski1906}, Onsager's regression hypothesis \cite{Onsager1931a,Onsager1931b}, and the linear response theory \cite{Kubo1957}.  The F-D relation also has practical importance. It allows deducing nonequilibrium dynamics from equilibrium measurements, and justifies the relation between macroscopic dynamics and microscopic level simulations, e.g., calculating the diffusion constant. In recent years, fluctuation theories of nonquilibrium processes, especially for systems far from equilibrium  received great attention \cite{Evans1993,Cugliandolo1994,Jarzynski1997,Crooks2000,Marconi2008}. 

 On studying problems from physics, chemistry, cellular biology, ecology, engineering, finance, and many other fields,  the following form of equations are widely used  \cite{vanKampen2007, Gillespie2000,Cobb1981},
\begin{eqnarray}
dx_i/dt = G_i(\mathbf{x})+\sum_{j=1}^m g_{ij}(\mathbf{x}) \zeta_j(t), i=1,\cdots, n. \label{eqn:stochaseqn}
\end{eqnarray}
In general $m$ and $n$ may be different,  $\zeta_j(t)$ are temporally uncorrelated, statistically independent Gaussian white noise with the averages satisfying $<\zeta_j (t) \zeta_j'(tÕ)> =  \delta_{jj'} \delta(t-tÕ)$, 
$\mathbf{g}(\mathbf{x})$ is related to the $n\times n$ diffusion matrix $\mathbf{gg}^T=2 \mathbf{D}/\beta$, where the superscript $T$ refers to transpose. For a physical system $\beta$ is the inverse of the Boltzmann's constant multiplying temperature. 
For a non-physical system, one can define an effective temperature relating to $\beta$.  
Recently we proved that  there exists a  mapping between a stochastic dissipative system described by Eqn. \ref{eqn:stochaseqn} and a thermostated Hamiltonian system \cite{Xing2009}.
 The mapping allows many results from equilibrium statistical physics directly applicable to nonequilibrium processes. Specifically in this work, we will derive  the F-D relation applicable to processes far from  equilibrium. An F-D relation, if exists, would allow predicting the relaxation dynamics to a steady-state through measurements of steady-state fluctuations. The latter are in general easier to measure.

\section{Theory}
\subsection{Existence of mapping between linear stochastic dissipative systems and Hamiltonian systems}
First we briefly summarize the main result of  \cite{Xing2009}. The mapping is based a seminal work of Ao, which shows that one can always construct a symmetric matrix $\mathbf{S}$ and an anti-symmetric one $\mathbf{T}$, and transform Eqn. \ref{eqn:stochaseqn} into, \cite{AoJPhys2004}
\begin{eqnarray}
(\mathbf{S}+\mathbf{T})\frac{d\mathbf{x}}{dt}  &=& (\mathbf{S}+\mathbf{T}) (\mathbf{G(x)} +\mathbf{g(x)}\zeta(t) ) \nonumber\\
&=&  - \nabla_\mathbf{x} \phi(\mathbf{x}) +\mathbf{g}'(\mathbf{x})\zeta(t) \label{eqn:transeqn}
\end{eqnarray}
where $\phi$ is a scalar function corresponding to the potential function in a Hamiltonian system satisfying 
$(\partial \times \partial\phi)_{ij}\equiv (\partial_i\partial_j-\partial_j\partial_i)\phi=0$, and
$\mathbf{g' g'}^T =2 \mathbf{S}/\beta$.  Then $\mathbf{S}$ and $\mathbf{T}$ are uniquely determined by 
\begin{eqnarray}
\partial\times[\mathbf{(MG(x)}]=0,(\mathbf{M})^{-1}+(\mathbf{M})^{-T}=2\mathbf{gg}^T, \label{eqn:M_eqn}
\end{eqnarray}
where $\mathbf{M=S+T}$, with proper choice of the boundary conditions. In \cite{Xing2009}, we first demonstrated that starting with the corresponding Fokker-Planck equations, one can derive the transformation matrix more transparently following a standard procedure used by Graham and by Eyink {\it et al.} previously \cite{Graham1977, Eyink1996}. 
Then we showed that one can map the dynamics described by Eqn. \ref{eqn:transeqn} 
to a  Hamiltonian system in the zero mass limit, 
\begin{eqnarray}
 H  = \frac{(\mathbf{\tilde{p}-A(x)})^2}{2m}+\phi(\mathbf{x}) \nonumber\\
 +  \sum_{\alpha=1}^{N_\alpha} \left[ 
         \sum_{j=1}^N \left( \frac{1}{2} p_{\alpha j}^2 
         + \frac{1}{2}\omega_{\alpha j}^2(q_{\alpha j}-a_{\alpha}(\mathbf{x})/(\sqrt{N} \omega_{\alpha j}^2))^2 
         \right) \right]
\end{eqnarray}
The term $\mathbf{A}$ is a vector potential satisfying $T_{ij} = \sum_j \left[\frac{\partial A_i}{\partial x_j} - \frac{\partial A_j}{\partial x_i}\right] $. The last term is a type of bath Hamiltonian discussed by Zwanzig \cite{Zwanzig1973}.  The Hamiltonian mathematically corresponds to Dirac's constrained Hamiltonian \cite{Dirac2001}. It describes a massless particle, coupled to a set of harmonic oscillators, moving in a hypothetical $n$-dimensional conservative scalar potential and magnetic (the vector potential) field. 
The steady state distribution is thus given by the Boltzmann distribution of the Hamiltonian system,
\begin{eqnarray}
\rho_{ss}(\mathbf{x}) = \frac { \int d\mathbf{p} d\mathbf{Y}  \exp(-\beta H )}
 { \int d\mathbf{x}d\mathbf{p} 
d\mathbf{Y}\exp(-\beta H )}\
=   \frac{\exp(-\beta \phi )}
 { \int d\mathbf{x} \exp(-\beta \phi )}\
\label{eqn:rho_ss}
\end{eqnarray}
where $\mathbf{Y}$ represents all the bath variables. Eqn.  \ref{eqn:rho_ss}  is also conjectured by Ao \cite{AoJPhys2004}, and a general proof is given in \cite{Yin2006}. 

\subsection{General fluctuation-dissipation theorem}

Let's denote the steady-state ensemble average of a generic dynamic quantity $\mathbf{O}$  as,
\begin{eqnarray}
<\mathbf{O}> =  \int d\mathbf{x}  \hspace{0.5pt} \mathbf{O}\rho_{ss}(\mathbf{x}) 
\end{eqnarray}
Consider a system initially at a steady state defined by Eqn. \ref{eqn:stochaseqn} with an extra infinitesimal perturbation $\delta \mathbf{G}(\lambda)$, where $\lambda$ refers to the parameters being perturbed. 
The corresponding perturbation Hamiltonian term is given  by   $\nabla (\delta H) = -\mathbf{(M'G'-MG)} \approx -\mathbf{M}\delta \mathbf{G} - \mathbf{\delta M}\mathbf{G}$, 
where $\mathbf{\delta M}$ is the variation of $\mathbf{M}$ due to the perturbation. 
At time 0 $\delta \mathbf{G}$  is removed, and the system relaxes to the steady-state of $\mathbf{G}$. Or  the Hamiltonian of the corresponding mapped system changes from $H'=H+\delta H=H+\frac{\partial \phi}{\partial\lambda}\cdot \delta \lambda$.  Then the nonequilibrium relaxation dynamics of $\mathbf{O}$ follows
\begin{eqnarray}
&&\bar{\mathbf{O}}(t) - <\mathbf{O}> = \frac{\int d\mathbf{x} d\mathbf{p} \hspace{0.5pt} \mathbf{O}(t)\exp(-\beta H')} 
 {\int d\mathbf{x} d\mathbf{p} \exp(-\beta H')} - <\mathbf{O}> \nonumber\\
 &&  \approx    -\beta [<\mathbf{O}(t)\frac{\partial \phi({\mathbf{x}(0)})}{\partial\lambda})>
    - <\mathbf{O}><\frac{\partial \phi}{\partial\lambda}>] \cdot \delta \lambda(0) \nonumber\\
    &&  \label{eqn:GFD} 
\end{eqnarray}
In the above derivation, we used the stationary property of equilibrium ensemble average, so $ <\mathbf{O}(t)> = <\mathbf{O}(0)> =<\mathbf{O}> $.
This is the generalized F-D relation for systems obeying or violating detailed balance, which states that nonequilibrium relaxation dynamics can be predicted from steady-state fluctuations. 
If one relates the relaxation function to the linear response function,  
 \begin{eqnarray}
\bar{O}_i(t) - <O_i>   \approx   \int_{t_0}^t dt'  \chi_{ij}(t-t') \delta \lambda_j(t')  \label{eqn:def_response} 
\end{eqnarray}
one obtains the differential form of the F-D relations,
\begin{eqnarray}
 \chi_{ij}(t-t')= -\beta \frac{d}{dt} \left[ <O_i(t)\frac{\partial \phi(\mathbf{x}(t'))}{\partial\lambda_j})> \right] \label{eqn:diff_GFD} 
\end{eqnarray}
Eqns \ref{eqn:GFD} and \ref{eqn:diff_GFD} are the central results of this work. These results are actually mathematically trivial with the replacement $\phi=-\ln\rho_{ss}$ \cite{Marconi2008}.  The mapping, however,  provides direct connection between $\rho_{ss}$ and the Langevin equations (see especially case 4 in the next section), and allows unified treatment for systems obeying or violating detailed balance.    One can generalize the results discussed in this work to higher orders of $\delta \mathbf{G}$. 
 
One type of perturbation of special interest is a system coordinate couples to some constant external force linearly, $\delta H = \mathbf{f\cdot x}$, which corresponds to $\delta \mathbf{G = - M^{-1}f}$ with $\delta\mathbf{M} = 0$. Notice that $\delta\mathbf{G}$ is in general a nonlinear function of $\mathbf{x}$.  This situation has been previously discussed by Graham, and by Eyink {\it et al.} \cite{Graham1977, Eyink1996}. 
Under this special type of perturbation, all the familiar results obtained on studying relaxations near an equilibrium state follow \cite{Kubo1991}. One can define  a response function $\chi(t,t')$ so $\bar{\mathbf{x}}(t)-<\mathbf{x}> =\int_{-\infty}^\infty dt' \mathbf{\chi}(t-t')\mathbf{f}(t') + O(\vert\vert f\vert\vert ^2)$. The function  $\chi(t-t')$  is stationary and satisfies the Kramers-Kr\"{o}nig relations. If $\mathbf{f}$ is time varying with a monochromic frequency,
 $\mathbf{f}=Re[\mathbf{f}_\omega \exp(i\omega t)+\mathbf{f}^*_\omega \exp(-i\omega t)] $, the system absorbs "energy", with the absorption spectrum $abs(\omega) \propto \omega^2\int_0^\infty dt  \mathbf{f}^T_\omega <(\mathbf{x}_t-<\mathbf{x}>(\mathbf{x}_0-<\mathbf{x}>)^T> \mathbf{f}^*_\omega \cos(\omega t)$ \cite{Kubo1991}.
%

\section{Special cases and numerical tests}

Here we will show that several versions of the previously derived generalized F-D relations are special cases of the above results. We will also consider several  pedagogical examples, especially biochemical networks,  to demonstrate the validity of our results. 

{\bf Case 1}: The first one is an analytically solvable irreversible linear chemical network (Fig. 1a). The system is initially at a state with $k_0 + \delta k$. 
Then at time 0 it changes to $k_0$, {\it i.e.}, the inflow flux varies. The system relaxes to a new steady-state.   The perturbation is $\delta \mathbf{G} = (\delta k,0,0)$, and the corresponding Hamiltonian perturbation term is $\delta H = -(x_1,x_2,x_3) \mathbf{M} \delta \mathbf{G}$. 
The dynamic equations are,
\begin{eqnarray}
\frac{d}{dt} 
        \left( \begin{array} {c} 
                               {x_1} \\ 
                               {x_2 }  \\ 
                               {x_3} 
          \end{array} \right) =
\left(\begin{array}{ccc}
-2 &    0   & 0 \\
1          & -1 & 1\\
1          &  0      & -1  
  \end{array} \right) 
\left(\begin{array}{c} {x_1} \\ {x_2} \\ {x_3} \end{array} \right) \nonumber\\
+ \left(\begin{array}{c} k_0  \\ 0  \\ 0  \end{array} \right)
+ \left(\begin{array}{c}  z \zeta_1(t) \\ z \zeta_2(t) \\ z\zeta_3(t) \end{array} \right)
\end{eqnarray}
For simplicity of the following discussions, we write the above equation in the form $d \mathbf{x}/dt =  \mathbf{K x + \mathbf{b}}+ z \zeta(t)$. The predicted relaxation function through the fluctuation-dissipation relation is given by,
\begin{eqnarray}
( \bar{x}_i(t) - <x_i>)_{FD} \propto \mathbf{C}\cdot  \mathbf{M}\cdot\delta \mathbf{G} \label{eqn:R_theory}
\end{eqnarray}  
where $(C_{ij}(t))  = <(x_i(t)-<x_i>)(x_j(0)-< x_j>)> $. We also tested an incomplete relaxation function by setting $\mathbf{M}$ as an identity matrix,
\begin{eqnarray}
( \bar{x}_i(t) - <x_i> )_{incomplete} \propto \mathbf{C}\cdot\delta \mathbf{G} \label{eqn:R_incomplete}
\end{eqnarray}  
Analytical solutions are given in the appendix. Fig. 1 shows that $\Delta x_i(t)$ calculated from the F-D relation (Eqn \ref{eqn:R_theory}) reproduces the exact result given by Eqn \ref{eqn:R_exact} (solid line). 
However, without the term $\mathbf{M}$, the incomplete F-D relation Eqn \ref{eqn:R_incomplete} predicts an initial increase of $x_2$ (and similar for $x_3$).
 Physically, spontaneous fluctuation of $x_1$ is anti-correlated with those of $x_2$ and $x_3$ concentrations through reactions 1 and 3. The incomplete F-D relation erroneously attributes this mechanism to the relaxation dynamics. 

In the second example, we still use the network shown in Fig. 1a, but  assume that all the reactions except reaction 0 follow the Michaelis-Menten kinetics, {\it e.g.}, 
\begin{eqnarray}
\frac{dx_1}{dt}= k_0-\frac{v_1 x_1}{K_1 +x_1}-\frac{v_3 x_1}{K_3+ x_1} + z\zeta_1(t)
\end{eqnarray} 
and similar expressions for $x_2$ and $x_3$.The relaxation functions are simulated numerically. Appendix B gives computational details.
 First the perturbation is again on $k_0$. Fig. 2 shows that the relaxation dynamics at small perturbations can be
well reproduced by the linear FD theory. Large deviation at larger perturbation amplitudes may be due to breakdown of the linear approximation for either
$\mathbf{G}$ or in Eqn. \ref{eqn:GFD}. Alternatively, we added perturbation on $v_1$. When $v_1$ decreases from an initial larger value, $x_1$ increases
due to decreased consumption through reaction 1; $x_2$ first decreases, then recovers to a new steady state value after the flux from $x_3$ flows in ; $x_3$ increases since 
the flux of the competing $x_1\rightarrow x_2$ pathway decreases, while the incoming flux to $x_1$ is unchanged. Fig. 3 shows that the FD theory reproduces all these
behaviors. We found deviation only at rather large perturbation values.  This example demonstrates that the linear approximation works for more realistic models.

{\bf case 2}: If one takes $O_i = \frac{\partial\phi}{\partial\lambda}_i$, Eqn. \ref{eqn:diff_GFD} gives the relation derived by Prost {\it et al.} from the Hatano-Sasa equality recently \cite{Prost2009}.

{\bf case 3}: Consider a particle moving along a one-dimensional periodic potential  $V(x+L)=V(x)$ and under a constant force $F$. The corresponding Langevin equation is  
\begin{eqnarray}
dx /dt= \partial_x V_{eff}+\sqrt{2D} \zeta(t) \label{eqn:dtheta}
\end{eqnarray}
with $V_{eff}=V+Fx$.
The "steady state" distribution (projected to and normalized within a range of $L$) is \cite{Reimann2002},
\begin{eqnarray}
\rho_{ss}(x) = \mathcal{N} \int_0^{L}dy e^{\left[\left( H_0(\sin(x+y)-\sin x)-F y \right)/D\right]}
\end{eqnarray}
where $\mathcal{N}$ is the normalization factor. Then Eqn. \ref{eqn:GFD} can be rewritten as,
\begin{eqnarray}
 \bar{O}(t)-<O> \approx \frac{\beta}{D} \left[ <O\frac{\partial V_{eff}(x)}{\partial \lambda}> - \right. \nonumber\\
  \left. <O><\frac{\partial V_{eff}(x)}{\partial \lambda}>  \right] 
 -  \frac{\beta}{D} \left[ <O B> -<O><B>  \right]
\end{eqnarray}
where $B= \int_0^{L}dy \frac{\partial V_{eff}(x+y)}{\partial\lambda}  e^{\left[\left( H_0(\sin(x+y)-\sin x)-F y \right)/D\right]}$. The second term in the right hand side of the above equation vanishes when $F=0$, and so the relation reduces to the F-D relation near equilibrium. In the special case $O$ and $\frac{\partial V_{eff}(x)}{\partial \lambda}$ are conjugate variables, and the motion is along a circular track, it remains to show that the current result is equivalent to what derived by Chetrite {\it et al.} \cite{Chetrite2008}, and experimentally tested by Solano {\it et al} \cite{Solano2009}.

{\bf case 4}: To further test the F-D model in nonlinear cases, we considered a model studied by Zhu {\it et al.} \cite{Zhu2006},
\begin{eqnarray}
\frac{d}{dt}\left(\begin{array}{c} x_1\\ x_2 \end{array} \right)
&=& -(\mathbf{S}+\mathbf{T})^{-1}\nabla\psi + \mathbf{\zeta} \\
\mathbf{S} &=& \frac{(q^2-1)^2}{(q^2-1)^2+1}\left(
\begin{array}{cc}
1 & 0 \\
0 & 1
\end{array}
\right) \nonumber\\
\mathbf{T} &=& \frac{(q-1+\epsilon) q^2}{(q^2-1)^2+1}\left(
\begin{array}{cc}
0 & -1 \\
1 & 0
\end{array}
\right) \nonumber\\
\phi &=& \frac{1}{2}(q-1)^2 = \frac{1}{2}(\sqrt{x_1^2+x_2^2}-1)^2 \nonumber\\
<\zeta_i\zeta_j> &=& 2D_{ij}/\beta
\end{eqnarray}
 In the last expression, the diffusion matrix $\mathbf{D}$ is given by the symmetric part of $(\mathbf{S+T})^{-1}$, and we used $\beta = 20$. Compared to what used by Zhu {\it et al.}, we added an extra term $\epsilon=0.01$ to avoid singularity at $q=1$ in the introduced perturbation, $\delta \mathbf{G} = (\mathbf{S+T})^{-1}\mathbf{f} $, even with elements of the constant vector $\vert f_i\vert<<1$. Results in Fig. 4a show that indeed with this special choice of the perturbation, $\Delta {x}_1(t)  \propto f_1<x_1(t)x_1(0)> +f_2 <x_1(t)x_2(0)>  $, (noticing $<x_i>=0$),  in the linear response regime. This result is striking given that 
 the perturbation $\delta \mathbf{G}$ is highly nonlinear. The response function shows damped oscillations. Therefore, one expects Van Vleck-Weisskopf-Fr\"{o}hlich type resonance absorption \cite{Kubo1991}. The resonance is confirmed in Fig. 4b. Here we talk about the mapping Hamiltonian. The mapping "energy" is not the same as the physical energy. Nevertheless, one expects that the resonance absorption may manifest itself in the dynamics of $\mathbf{x}$ under periodic perturbation. 
 We added a perturbation energy term $0.005\cos(\omega t)(x_1+x_2)$. Our stochastic simulations did find that the amplitude of the $<x_i>$ oscillation driven by the perturbation shows a maximum near that suggested by the absorption spectrum
  (see Fig. 4c).    
  
\section{Discussions and concluding remarks}
In this work, we showed how a recently discovered mapping between dissipative and Hamiltonian systems can lead to a general F-D relations. 
Further studies are needed for potential applications of the generalized F-D relation on studying systems described by Eqn \ref{eqn:stochaseqn} \cite{Marconi2008,Stratonovich1993}. As an example, one can use it on biological network reconstruction. The matrix $\mathbf{M}$ contains information on the network topology and parameter values. Researchers proposed to use the relaxation dynamics to derive information of a biological network, especially a linear network \cite{Arkin1997,Gardner2003}. A potential problem is that the relaxation dynamics alone may contain insufficient information to resolve a network topology and parameters, and to distinguish competing models. The F-D relation provides additional constraints, and may serve as criteria for evaluating different models. It remains to be examined if one can distinguish abnormal (e.g. cancer) cells from normal cells based on their differences of relaxation and fluctuation dynamics.

 We suggest that one should further exploit the analogy between the dissipative and Hamiltonian systems. Many existing results for the latter may lead to new understanding for the former, provided appropriate quantities can be identified. As an example, researchers have used oscillating signals to perturb biological systems. They found that the oscillation amplitude of the variable being driven shows a maximum at certain driving frequency \cite{Lipan2005,Mettetal2008}. Here we showed that this phenomenon can be interrpreted as the Van Vleck-Weisskopf-Fr\"{o}hlich type resonance absorption \cite{Kubo1991}. The analogy may suggest further technical development in biological network studies parallel to the linear and nonlinear spectral methods used in physics and chemistry.  Study analogous to Onsager's reciprocal relations is another example \cite{Onsager1931a, Onsager1931b}.  A potential difficulty is to relate the mapped quantities to physically measurable quantities. 

The relaxation dynamics is important for characterizing the dynamics of a system. For example, the resistor model suggests that a HIV slows down its relaxation from an unstable active excited state to a stable dormant state through a series of intermediate steps in its regulation network \cite{Weinberger2007}. 
Wang and coworkers found that many biological networks have rugged potential profiles (as defined by Eqn. \ref{eqn:rho_ss}), which resemble what observed in protein folding studies \cite{Wang2006,Lapidus2008, Wang2008}. 
 The finding reinforces that dynamics of biological networks is complex.
Implications of noises in biological systems have been extensively discussed \cite{Rao2002}. Here the mapping relates noise intensity to an effective temperature. Consequently
we suggest that noise prevents a system from trapping into large numbers of undesired intermediate steady-states a complex system may have. Therefore, counter-intuitively existence of proper amount of noises, together with funneled landscapes \cite{Wang2006}, enhances rather than destroy the robustness of a network. Glass transition is an active research topic for the relaxation process of a physical system to an equilibrium state \cite{Kauzmann1948, Iben1989,Fischer1993}. 
A typical biological network is highly inhomogeneous, and is full of competing interactions. Thus its dynamics in some sense resembles that of spin glasses but with nonrandom quenching \cite{Fischer1993}. 
It is interesting to notice that the mapping between a dissipative and a Hamiltonian system suggest possible glassy behaviors for relaxation to a nonequilibrium steady-state. If being identified, (noticing glassy state has been identified  in protein folding studies \cite{Iben1989,Pande2000}, also refer chapter 6 of \cite{Fischer1993} for a spin glass model in the form of Eqn. \ref{eqn:stochaseqn}), it may have profound practical implications. For example, a patient might then have prolonged recovery period to the normal homeostatic state after a disease. One solution is to increase the temperature ($\it{i.e.}$, noise intensity in the related regulation network) above the critical glass transition temperature. Indeed cells can regulate their noise levels \cite{Raser2005}.

In this work we only established a framework.  Most of our discussions focus on biological examples. They equally apply to problems in other fields sharing the same mathematical structure, e.g., a stock market model described by Eqn. \ref{eqn:stochaseqn}.




 
\section{Appendix}

\subsection{A}
Here we give the analytical solutions of the  first example in case 1.
The formal solution of the stochastic differential equations is,
\begin{eqnarray}
\mathbf{x(t)} & =& <\mathbf{x}>+ \exp{(\mathbf{K} t )} (\mathbf{x(0)-<x>}) \nonumber\\
&& + z \int_0^t  \exp{ (\mathbf{K} (t-\tau))} \zeta (\tau) d\tau 
\end{eqnarray}
where $<\mathbf{x}>$ is the solution of $\mathbf{K\cdot x+b}=0$.
The spontaneous fluctuation correlation functions are given by (with $\beta=1$),
\begin{eqnarray}
&&(C_{ij}(t))  = <(x_i(t)-<x_i>)(x_j(0)-< x_j>)> \nonumber\\
&& \propto \int d\mathbf{x(0)} \exp[-(\mathbf{x(0)-<x>})^T\cdot
\frac{1}{2} \mathbf{M\cdot K}\cdot (\mathbf{x(0)} \nonumber \\
  &&-<\mathbf{x}>)] 
   \left( \exp{(\mathbf{K} t )} (\mathbf{x(0)-<x>})\right)_{i} (x_j(0)-< x_j>) \nonumber \\
 && \propto \sum_k \exp {(\mathbf{K} t)}_{ik} \left[
   \left(\frac{1}{2} \mathbf{M\cdot K} \right)^{-1}_{kj}  (1+ \delta_{kj} ) \right] 
\end{eqnarray}
The relaxation functions for the system relaxing from a steady state corresponding to $k_0+\delta k$ to that of $k_0$ are given by,
\begin{eqnarray}
&& \Delta x_i(t) = \bar{x}_i(t) - <x_i> \propto \nonumber\\
 && \int d\mathbf{x(0)} \exp[-(\mathbf{x(0)-<x'>})^T\cdot \frac{1}{2}
\mathbf{M\cdot K}\cdot \nonumber\\
&& (\mathbf{x(0)-<x'>})] 
 \left( \exp{(\mathbf{K} t )} (\mathbf{x(0)-<x>})\right)_{i} \nonumber\\
 && \propto   \left( \exp{(\mathbf{K} t )} (\mathbf{<x'>-<x>})\right)_{i} \label{eqn:R_exact}
\end{eqnarray}
where $<\mathbf{x'}>$ satisfies $\mathbf{K\cdot <x'>}+{(k_0+\delta k,0,0)}^T =0$.

We calculate the transformation matrix $\mathbf{M}$ using the result of Kwon {\it et al.} \cite{Kwon2005}. For the choice of parameter values in this work, expression of the transformation matrix is given by,   
\begin{eqnarray}
 \mathbf{M} =
 \frac{1}{81z+50z^3}
 \left(\begin{array}{ccc}
81+25z^2 & -36+15z & -27-20z\\
36+15z  & 81+9z^2 & 45-12z\\
27-20z & -45 - 12z& 81 + 16z^2\nonumber
\end{array}
\right)
\end{eqnarray}

we want to point out that it is well known that the Langevin-type equations are only mathematical approximations of the physical processes, which may lead to unphysical results. For example, the concentration of a specie in a network may go to negative values. In the above derivations we assume  that the probability of obtaining zero and further negative concentrations is negligible, and thus we can perform the integration from $-\infty$ to $\infty$.

\subsection{B}
For example 2 in case 1,
the Langevin equations are propagated by 
\begin{equation}
x_i(t_N) = x_i(t_{N-1}) + \Delta t G_i(\mathbf{x}) + \sqrt{2 z \Delta t/\beta} \zeta_i(t),
\end{equation}
where $\zeta_i(t)$ is generated from a Gaussian distribution with zero mean and unit variance, and we have set $\beta = 1$ throughout the work. We used $\Delta t = 0.005$ in all calculations. To calculate the relaxation function, we first propagated the network with a perturbation ( $k_0+\delta k$ for results in Fig.2, and $v_1+\delta v_1$ in Fig. 3)  $4\times 10^3$ steps, so the system reaches a steady-state corresponding to the perturbed $k_0$. Then we switch back to the u.nperturbed system, set time 0, record the values of $x_i$ as $x_i(0)$, propagate for another $4\times 10^3$ steps. We calculated $\bar{x}_i(t)$  by averaging over $5\times 10^5$ such trajectories.  
The system is expanded linearly around the steady-state of the corresponding deterministic model. 
The predicted relaxation functions in Fig. 2 and 3 are calculated with $\delta k = 0.1$, and $\delta v_1 = 0.1$, respectively. 
 We found that the system dynamics is within linear response regime when the predicted relaxation dynamics agrees with the simulated ones. 

For the example in case 4, the Langevin equations are propagated similarly. Results are averaged over $5\times 10^5$ trajectories. Rigorously speaking, there is a correction term to the stochastic Langevin equation. Our numerical algorithm adopts the Itoh interpretation. In \cite{Xing2009}, we discussed the relation between the Itoh interpretation and the zero-mass interpretation adopted for deriving the mapping Hamiltonian. The expression of the correction term can be derived straightforwardly. However, the correction term is proportional to $1/\beta$, and is small for this system. Therefore we neglected this term here.

\subsection{C}
Here we prove that there is no need to consider $\frac{\partial \mathcal{N}}{\partial \lambda}$. 
Let's express $\rho_{ss}=\mathcal{N} \rho_0$, and $\phi(x)=-\frac{1}{\beta}(\ln\rho_0 + \ln\mathcal{N}+\phi_0)$, then,
\begin{eqnarray}
  &&-\beta [<\mathbf{X}(t)\frac{\partial \phi({\mathbf{x}(0)})}{\partial\lambda})>
    - <\mathbf{X}><\frac{\partial \phi}{\partial\lambda}>] \nonumber\\
 &&  =   <\mathbf{X}(t)\frac{\partial \ln\rho_0({\mathbf{x}(0)})}{\partial\lambda})>
    - <\mathbf{X}><\frac{\partial \ln\rho_0}{\partial\lambda}>] \nonumber\\
\end{eqnarray}




We thank  Kenneth S Kim for discussions.

\begin{figure}
\includegraphics[width=87mm]{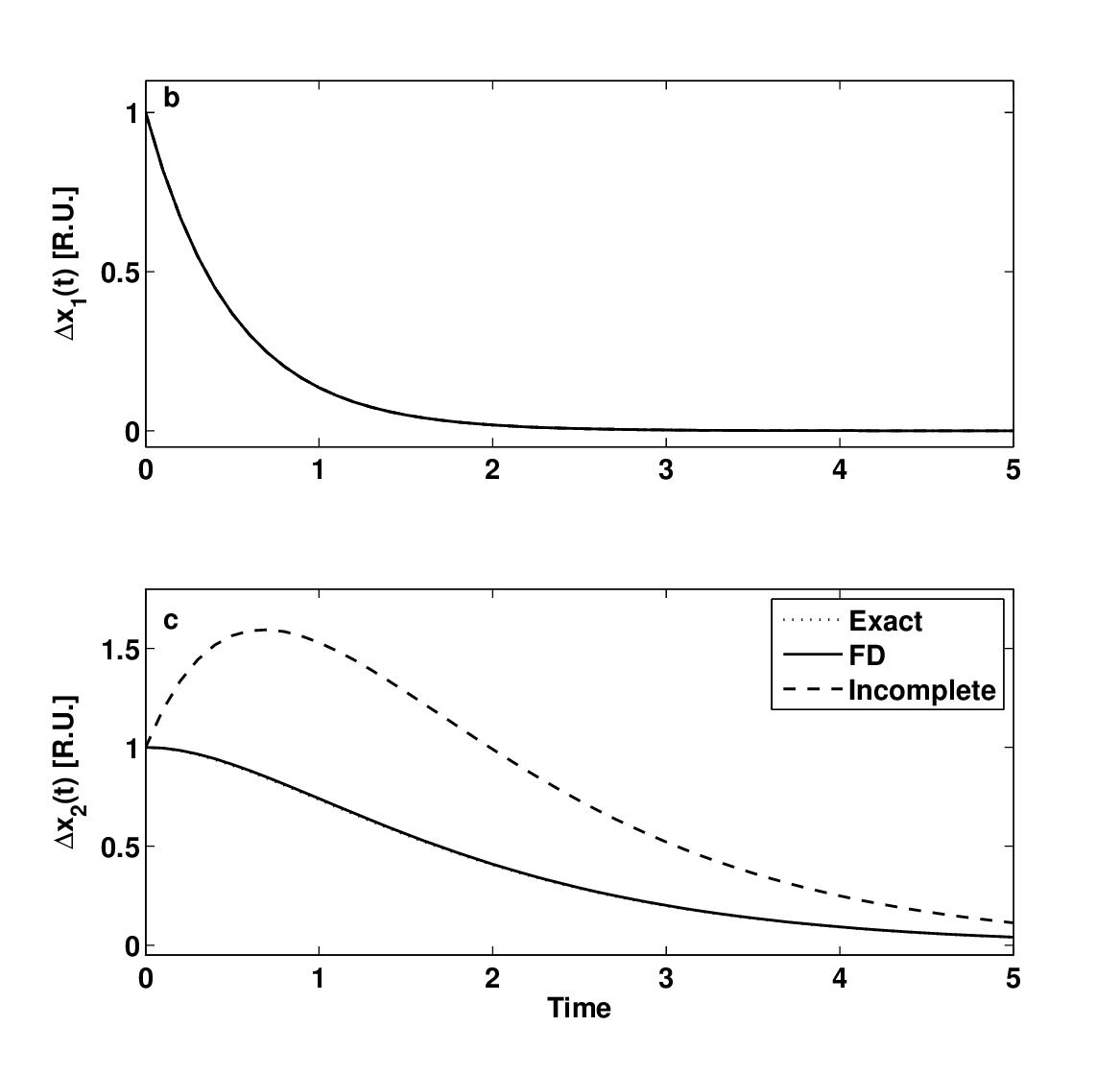}
\caption{Numerical tests with a linear network. (a) The network being tested. $k_0=k_1=k_2=k_3=k_4=1$.
(b) The normalized relaxation function  $\Delta x_1(t)=\left( \bar{x}_1(t)-<x_1> \right)/(\bar{x}_1(0)-<x_1>)$. (c) The  $\Delta x_2(t)$ calculated with Eqn \ref{eqn:R_theory} overlap with the exact result from Eqn \ref{eqn:R_exact} (solid line), but that calculated from Eqn \ref{eqn:R_incomplete} with $\mathbf{M}= 1$ gives wrong prediction (dashed line). Similar results for $\Delta x_3(t)$.         }\label{fig:linear}
\end{figure}

\begin{figure}
\includegraphics[width=87mm]{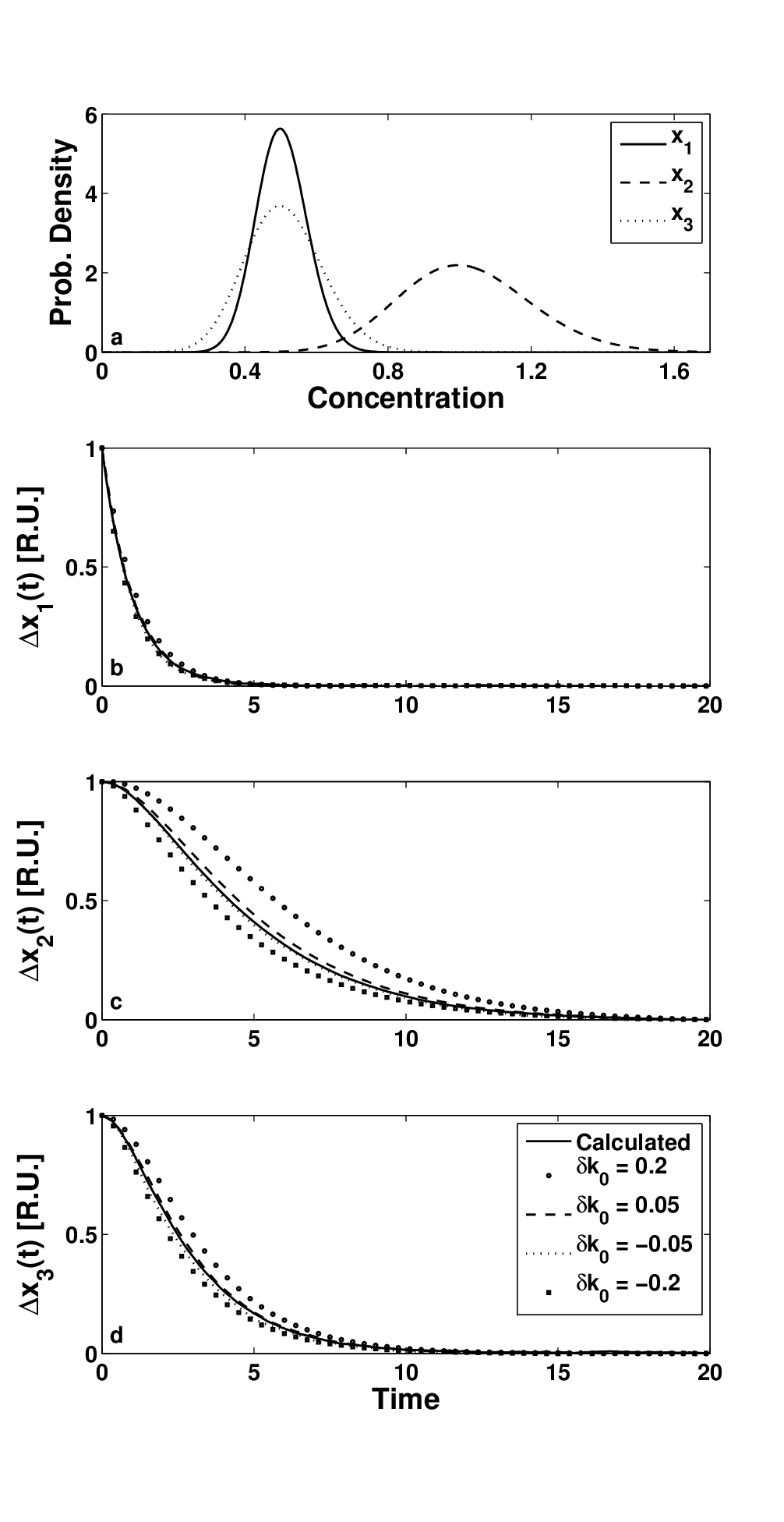}\label{fig:MM1}
\caption{Reaction network under Michaelis-Menten kinetics. Model parameters $k_0 = 1, v_i=1,K_i=0.5 \hspace{0.1cm} (i=1 - 4), \mbox{except } v_2=1.5, z=0.005$. (a): The steady-state distribution for the unperturbed system.  (b)-(d): The relaxation function with various perturbation $\delta k$. To compare results with different $\delta k$, all results are normalized by the maximum of $\vert \Delta x_t(t) \vert$.
  } 
\end{figure}

\begin{figure}
\label{fig:MM2}
\includegraphics[width=87mm]{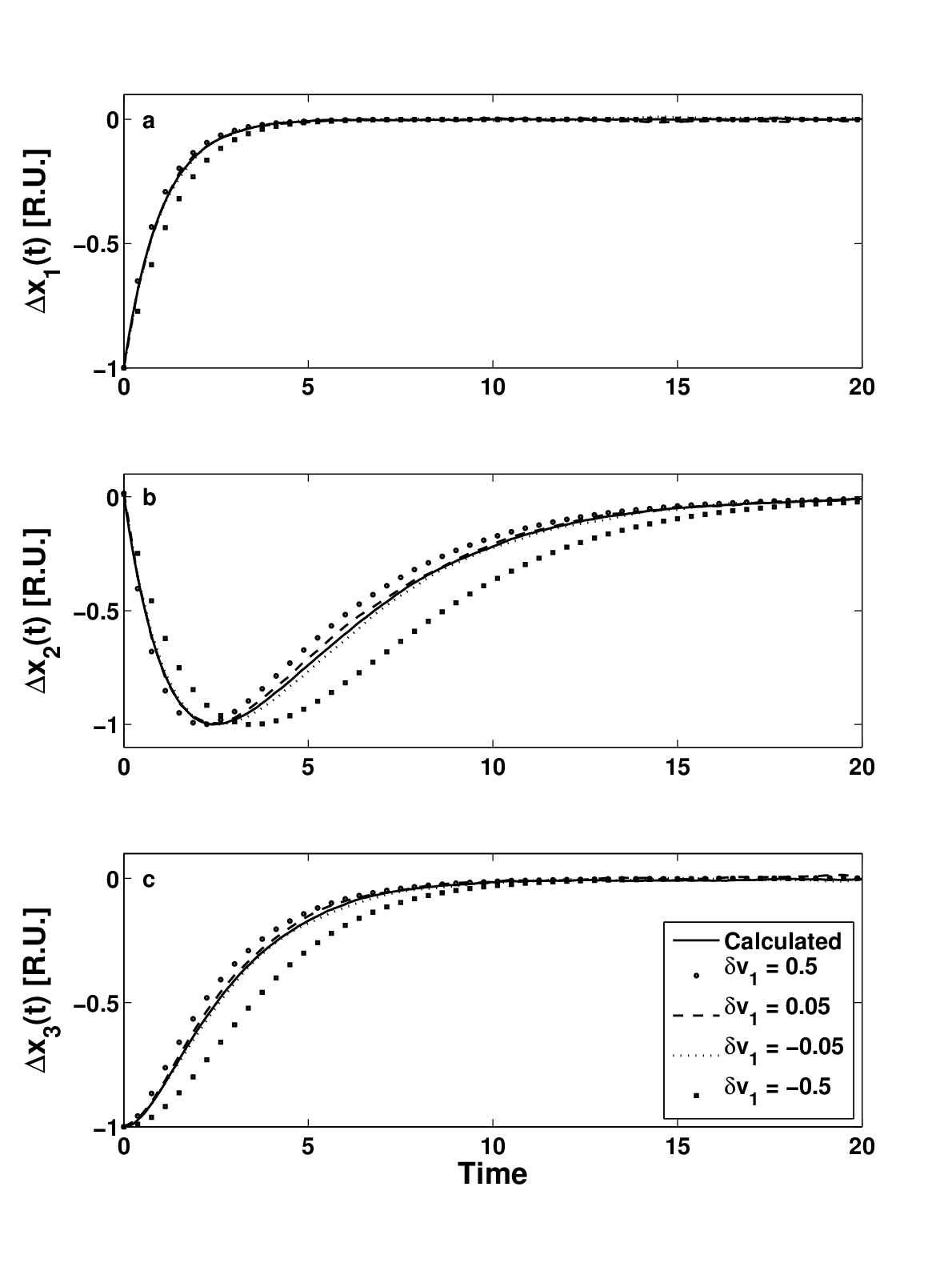}
\caption{Same as Fig. 2, except the perturbation is on $v_1$.      }
\end{figure}

\begin{figure}
\label{fig:LC}
\includegraphics[width=87mm]{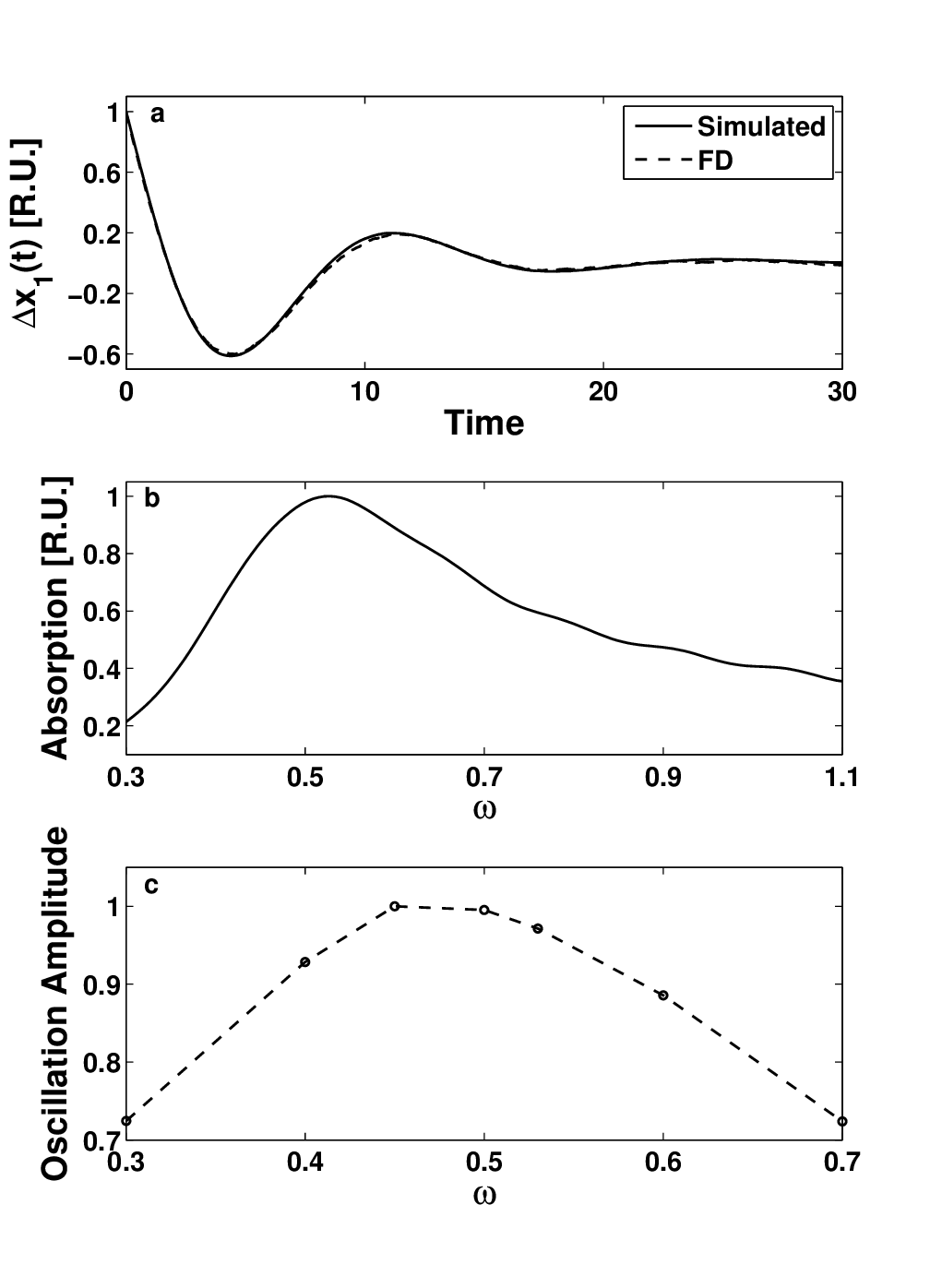}
\caption{Numerical tests with a model showing limiting cycle dynamics. 
(a) The normalized relaxation function  $\Delta x_1(t)=\left( \bar{x}_1(t)-<x_1> \right)/(\bar{x}_1(0)-<x_1>)$ from simulation and prediction of the F-D relation. $f_1=f_2=0.005$.
(b) The "absorption spectrum" calculated from the fluctuation correlation functions with $f_1=f_2=0.005\cos(\omega t)$
(c) The simulated amplitude of the $<x_i>$ oscillation driven by an oscillating perturbation used in b. }
\end{figure}








\end{document}